# Regrowth-free AlGaInAs MQW polarization controller integrated with sidewall grating DFB laser

Xiao Sun[1*], Song Liang[2], Weiqing Cheng[1], Shengwei Ye[1], Yiming Sun[1], Yongguang Huang[2], Ruikang Zhang[2], Jichuan Xiong[3], Xuefeng Liu[3], John H. Marsh[1], Lianping Hou[1]

[1] *James Watt School of Engineering, University of Glasgow, Glasgow G12 8QQ, U.K.*
[2] *College of Physics, Jilin University, Changchun 130012, China*
[2] *Institute of Semiconductors, Chinese Academy of Sciences, No. A35, East Qinghua Road, Haidian District, Beijing 100083, China*
[3] *School of Electronic and Optical Engineering, Nanjing University of Science and Technology, Nanjing 210094, China*
*Corresponding author: x.sun.2@research.gla.ac.uk*

**We report an AlGaInAs multiple-quantum-well integrated source of polarization controlled light consisting of a polarization mode converter(PMC), differential phase shifter(DPS), and a side wall grating distributed-feedback (DFB) laser. We demonstrate an asymmetrical stepped-height ridge waveguide PMC to realize TE to TM polarization conversion and a symmetrical straight waveguide DPS to enable polarization rotation from approximately counterclockwise circular polarization to linear polarization. Based on the identical epitaxial layer scheme, all of the PMC, DPS, and DFB laser can be integrated monolithically using only a single step of metalorganic vapor-phase epitaxy and two steps of III-V material dry etching. For the DFB-PMC device, a high TE to TM polarization conversion efficiency (98%) over a wide range of DFB injection currents is reported at 1555 nm wavelength. For the DFB-PMC-DPS device, a π/3 rotation of the Stokes vector was obtained on the Poincaré sphere with a range of bias voltage from 0 V to –4.0 V at $I_{DFB}$ =170 mA.**

## 1. Introduction

The ability to control the polarization state is of substantial interest in optical communication systems such as those using high-speed complex digital signal processing (DSP) to manipulate the TE-TM polarization state of light [1] or Stokes vector modulation and direct detection (SVM/DD) systems [2]. For those applications, there has been a growing interest in integrating a polarization controller with the light source, the detectors, or other components in a photonic integrated circuit (PIC). As an increasing number of devices such as laser diodes (LDs) [3] and electro-absorption modulators [4] utilize multiple-quantum-well (MQW) structures as the active region, it is desirable to design a polarization controller compatible with MQW structures.

Several different material systems and designs have been proposed for polarization controllers including the silicon on insulator (SOI) platform [5], the InGaAlAs/InP platform [4], and the InGaAsP/InP platform [6, 7]. A typical polarization convertor device comprises a cascade of polarization mode converters (PMCs) and MQW polarization-dependent phase shifters (PD-PSs) [4, 8] to achieve an arbitrary state of polarization (SOP). The light source can be an external LD or a monolithically integrated laser [3, 9]. However, many reported PMCs utilize bulk material as the core layer in the waveguide to achieve a high polarization conversion efficiency (PCE) in a short waveguide length. This approach requires relatively complicated regrown butt-joint PIC techniques to integrate a bulk PMC with an MQW-based PD-PS. In [8], a passive bulk PMC was monolithically integrated with an active MQW-based PD-PS by using the butt-joint technique and was a demonstrator of an efficient polarization controller in the InGaAsP/InP material system, however the device had no monolithic LD. In [3], an MQW-based PMC was monolithically integrated with an MQW-based FP LD by using the identical epitaxial layer (IEL) integration scheme, but the PCE was limited, being only 80% in [3, 10] and around 50% to 68% in [9]. The crucial issue for integrating PMCs with MQW devices is the inherent birefringence of the MQW, which disturbs the optimal rotation of the SOP. The two main mechanisms of SOP conversion are the mode-evolution method [11], utilizing the shifting of the propagating mode inside the waveguide, and the mode coupling method [7], which exploits the beating between two eigenmodes to enable polarization rotation along the PMC waveguide. Because mode coupling PMCs enable polarization conversion within a minimal waveguide length, it is important to reduce the absorption loss caused by the inter-band exciton transitions inside a MQW-based PMC monolithically integrated with a MQW-based LD [12]. To realize good mode-matching between the waveguides of monolithic PMCs and LDs, we have proposed an

AlGaInAs/InP MQW-based sidewall grating (SWG) distributed feedback (DFB) laser monolithically integrated with a stepped height waveguide PMC [13]. A novel MQW-based epitaxial structure was designed and optimized for both the PMC and SWG DFB laser through a series of full-wave simulations.

In this work, based on our simulation work in [13], for the first time an AlGaInAs MQW SWG DFB laser was fabricated and monolithically integrated with both a stepped waveguide PMC and differential phase shifter (DPS) based on the IEL integration scheme. This approach needs only a single step of metalorganic vapor-phase epitaxy (MOVPE) and two steps of III-V material dry etching. The monolithic DFB-PMC and DFB-PMC-DPS devices reported here avoid the complicated etch and regrowth processes required for conventional buried grating DFB laser structures and also the time-consuming butt-joint or selective area growth (SAG) PIC technologies. The DFB-PMC device has a high TE to TM PCE (98%) over a wide range of DFB injection currents ($I_{DFB}$) at an operating wavelength of 1555 nm. For the DFB-PMC-DPS device, a $\pi/3$ rotation of the Stocks vector (*SV*) was obtained on the Poincaré sphere with a range of bias voltage from 0 V to –4.0 V and $I_{DFB}$ =170 mA.

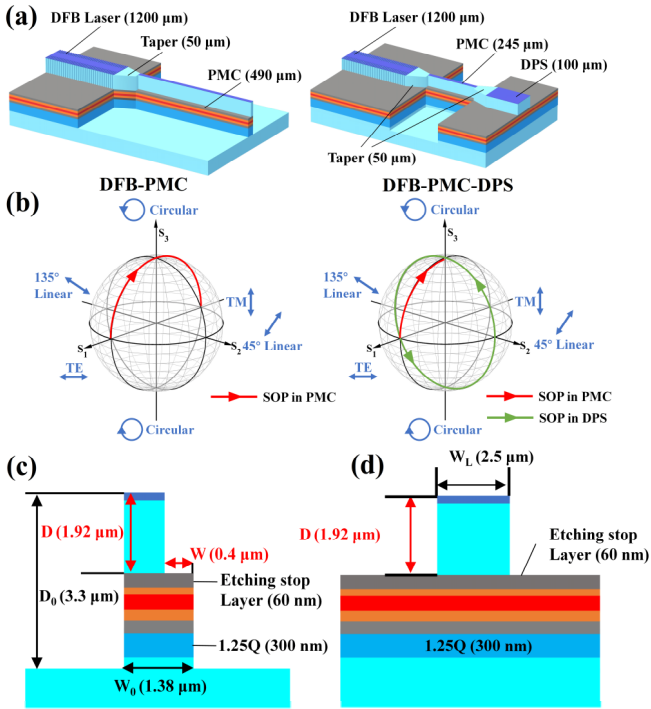

Fig. 1. (a) Schematic of the monolithic DFB-PMC device (left), and DFB-PMC-DPS device (right), (b) *SV* propagates inside DFB-PMC (left), and DFB-PMC-DPS device (right), (c) cross-section structure of the PMC, (d) cross-section structure of the DPS.

## 2. Device design and fabrication

The wafer structure used for the DFB-PMC and DFB-PMC-DPS is the same as that described in [13]. The wafer was grown on an InP substrate by metalorganic vapor-phase epitaxy (MOVPE). The room temperature PL peak of the QWs was located at a wavelength of 1530 nm. A 300-nm-thick 1.25Q InGaAsP (1.25Q means the PL wavelength of this material is 1.25 µm) layer is embedded below the MQW layer to increase the difference between the propagation constants of the two fundamental transverse modes ($TE_0$ and $TM_0$) of the PMC to reduce the half-beat length ($L_\pi$) and increase the PCE.

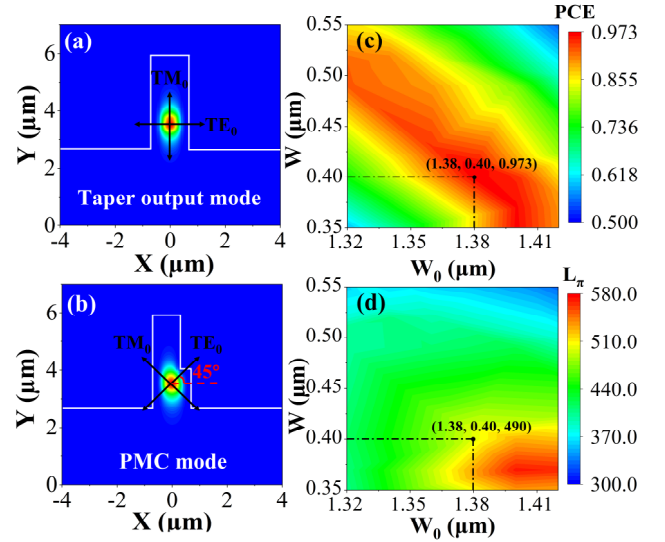

Fig. 2. (a)-(b) The fundamental eigenmodes in taper output (a) and PMC stepped-height ridge waveguide (b), (c)-(d) Calculated maximum PCE (c), and corresponding $L_\pi$ (d) as a function of waveguide width ($W_0$) and corner width ($W$).

Two kinds of polarization controllers – DFB-PMC and DFB-PMC-DPS – are proposed and are shown in Fig 1(a). Both devices contain the same 1200 µm long SWG ridge DFB laser. The ridge waveguide of DFB is 2.5 µm wide and 1.92 µm high. The gratings are of first order with a 50% duty cycle and formed by etching 0.6 µm recesses into the sidewalls of the waveguide, as shown in Fig.1(a). For the DFB-PMC device, LDs with three different Bragg wavelength were designed and fabricated. The grating period is 238 nm for a 1550 nm Bragg wavelength, 236.5 nm for a 1540 nm Bragg wavelength, and 240.3 nm for a 1565 nm Bragg wavelength. A quarter wavelength shift section was inserted at the center of the DFB laser cavity to ensure single longitudinal mode (SLM) oscillation. The length of the PMC in the DFB-PMC and DFB-PMC-DPS are 490 µm and 245 µm, respectively. The PMC is connected with the DFB LD using one 50 µm long taper for the DFB-PMC device, and is connected with the DFB laser and DPS respectively by two 50 µm long tapers for the DFB-PMC-DPS device. The simulated reflection between the shallow etched DFB and deep etched taper sections is about $7\times10^{-6}$, which has a negligible effect on the DFB performance. The simulated taper excess optical loss is 1%, i.e., 0.044 dB, which includes the scattering and mode mismatch loss, and can also be neglected. Fig.1(b) presents the *SV* rotation inside those two devices. For the DFB-PMC device, the *SV* rotates around the $S_2$ axis arriving on the $S_1$-$S_2$ plane, and the output mode becomes TM-polarized. For the DFB-PMC-DPS device, after the PMC, the *SV* rotates around the $S_2$ axis (red line) and arrives near the north pole, i.e., counterclockwise circular polarization, and then the *SV* is rotated around the $S_1$ axis within the $S_2$-$S_3$ plane by using the reverse biased DPS section (green line). Assuming a TE polarized input ($S$ = (1,0,0)), the variation of $S= (S_1, S_2, S_3)^{tr}$ and the PCE inside the PMC waveguide can be defined as [14, 15]:

$$S = \begin{pmatrix} S_1 \\ S_2 \\ S_3 \end{pmatrix} = \begin{pmatrix} 1 - 2\sin^2(2\varphi)\sin^2\left(\pi \dfrac{L_{PMC}}{2L_\pi}\right) \\ \sin(4\varphi)\sin\left(\pi \dfrac{L_{PMC}}{2L_\pi}\right) \\ \sin(2\varphi)\cos\left(\pi \dfrac{L_{PMC}}{L_\pi}\right) \end{pmatrix} \quad (1)$$

$$PCE = \sin^2(2\varphi)\sin^2\left(\pi \dfrac{L_{PMC}}{2L_\pi}\right) = \dfrac{1-S_1}{2} \times 100\% \quad (2)$$

where $\varphi$ is the rotated angle of the eigenmodes in the PMC waveguide, and $L_{PMC}$ is the length of PMC and in this work $L_{PMC}$ is equal to the half-beat length $L_\pi = \pi/(\beta_1-\beta_2)$ (where $\beta_1$ and $\beta_2$ are the propagation constants of the $TE_0$ and $TM_0$ eigenmodes in the PMC waveguide). To realize the *SV* rotation in Fig 1(b), the value of $\varphi$ should be 45° and $L_{PMC}$ is equal to $L_\pi$ for the DFB-PMC and $L_\pi/2$ for the DFB-PMC-DPS. The dimensions of the stepped-height PMC waveguide chosen here are $W_0$ =1.38 µm, $W$ = 0.4 µm, $D_0$ = 3.3 µm and $D$ =1.92 µm as shown in Fig 1(c). $D$ can be precisely controlled because the top 60 nm thick AlGaInAs waveguide layer acts as a dry etch stop layer when using a $CH_4/H_2/O_2$ ICP recipe. Fig. 1(d) presents the cross-section of the DPS, which is a symmetric and shallow etched ridge waveguide with a width of 2.5 µm and height of 1.92 µm, the same as that of the DFB section. Fig. 2(a) and (b) present the fundamental modes in the taper output section and PMC waveguide. The PMC eigenmode is optimized to rotate the electric/magnetic fields through 45°. After propagating a half-beat length $L_\pi$, the $S_2$ is rotated 180° around the $S_2$ axis defined from Eq. (1), and the output becomes TM-polarized. Hence to optimize the PCE and $L_\pi$ of the PMC waveguide, a full-wave simulation was made using an FDTD software package. The input light wavelength was set at 1550 nm. The calculated effective modal indexes ($N_{eff}$) of the fundamental TE and TM modes are 3.21109 and 3.20951 respectively. Fig.2 (c) and (d) show a contour plot of the calculated PCE and $L_\pi$ as a function of $W_0$ and $W$. The final optimum widths of the PMC waveguide chosen here are $W_0$ =1.38 µm and $W$ = 0.4 µm, which provide a high PCE (97.3%) and short $L_\pi$ (490 µm).

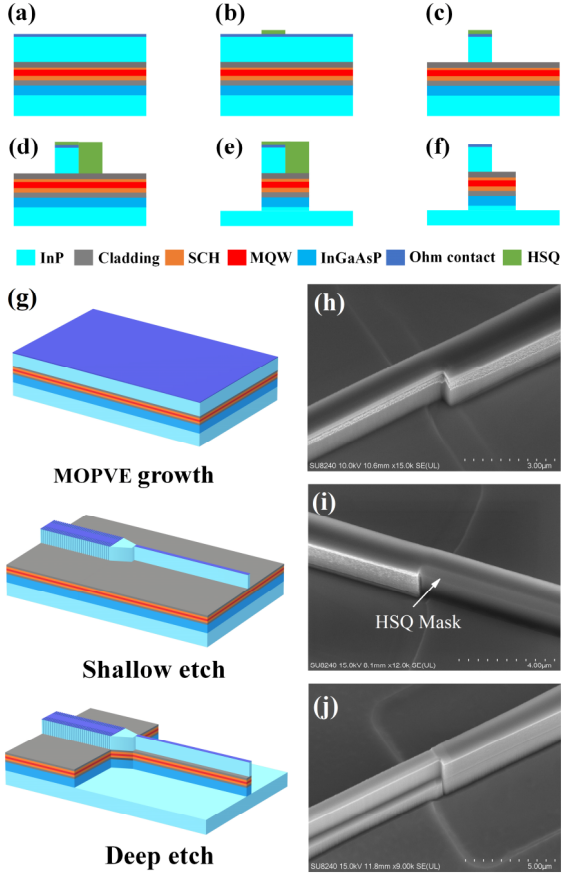

Fig. 3. (a)-(f) Fabrication procedures: (a) MOVPE epilayer growth, (b) EBL to define the laser and PMC first step waveguide, (c) ICP shallow etching, (d) EBL to define the second step waveguide of PMC, (e) ICP deep etching, (f) HSQ elimination, (g) workflow of monolithic DFB-PMC device fabrication, (h)-(j) SEM image of (h) after first shallow etch ,(i) second step EBL using HSQ photoresist, (j)PMC deep etch and HSQ elimination.

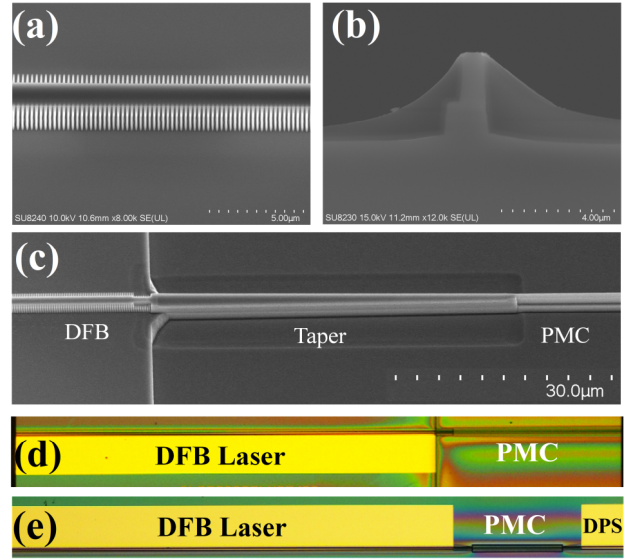

Fig. 4. (a)-(c)SEM image of (a)DFB laser with sidewall gratings, (b) output facet of PMC, (c) DFB-PMC device. (d)-(e) Microscope picture of the (d) DFB -PMC device and (e) DFB -PMC-DPS device.

The DFB-PMC device fabrication processes are presented in Fig.3. The wafer was grown on an InP substrate by MOVPE (Fig.3(a)). The DFB grating and PMC first step waveguide pattern were defined by e-beam lithography (EBL) and negative tone Hydrogen Silsesquioxane (HSQ) acted as an EBL resist and a hard mask for inductively coupled plasma (ICP) dry etching as shown in Fig. 3(b). In Fig. 3(c), the ridge was firstly etched to 1.89 µm in an ICP dry etch tool by a $Cl_2/CH_4/H_2/Ar$ gas mixture, the etch rate for InP, InGaAsP, and AlGaInAs being about 183 nm/min. Then, the gas recipe was changed to $CH_4/H_2/O_2$ to etch the ridge to 1.92 µm with an etch rate for InP/ InGaAsP of about 78 nm/min, and for AlGaInAs of 3 nm/s, achieving 26-fold selectivity. After the shallow etch, both the DFB section and PMC first step waveguide were protected by HSQ, and the second PMC step was defined by EBL as shown in Fig.3(d). A second stage of $Cl_2/CH_4/H_2/Ar$ ICP etching was used to etch the PMC ridge waveguide to a depth of 3.3 µm (Fig. 3(e)). Finally, all HSQ was removed by HF acid as shown in Fig. 3(f). The fabrication workflow is depicted in Fig. 3(g); only a single step of MOPVE and two steps of III-V material dry etching are required for the whole integrated device. Scanning electron microscope (SEM) images of the PMC waveguide after the first shallow etch, second step EBL and deep etch are shown in Fig.3(h), (i), and (j) respectively. The subsequent deposition of $SiO_2$ and HSQ passivation layers, $SiO_2$ window opening, P-contact deposition, substrate thinning, and N-contact deposition were the same as for conventional LD fabrication and can be referred to [16]. SEM images of the DFB grating, output facet of PMC, and DFB-PMC device are presented in Fig. 4(a)-(c). The optical microscope pictures of the completed DFB-PMC and DFB-PMC-DPS devices are depicted in Fig. 4(d) and (e) respectively. Finally, the devices were mounted epilayer up on a copper heat sink on a Peltier cooler. The heat sink temperature was set at 20°C and the devices were tested under CW conditions.

**Table 1. Measured parameters of the three DFB-PMC devices with different designed Bragg wavelengths**

| Designed Brag grating wavelength (nm) | SMSR (dB) DFB facet | SMSR (dB) PMC facet | Measured current (mA) and wavelength ranges (nm) | ACWRC (nm/mA) |
|---|---|---|---|---|
| 1540 | 27 | 8 | Current: 118 - 170<br>Wavelength: 1542.9-1544.2 | 0.0250 |
| 1550 | 29 | 12 | Current: 104 - 220<br>Wavelength: 1554.2-1556.6 | 0.0207 |
| 1565 | 38 | 20 | Current: 97 - 211<br>Wavelength: 1566.3-1569.4 | 0.0271 |

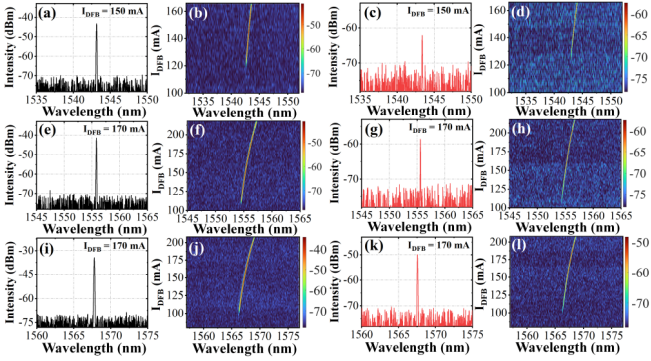

Fig. 5. (a)-(d) Measured optical spectrum for 1542 nm DFB-PMC device from DFB(a)(b) and PMC facet(c)(d), (e)-(h) optical for 1555 nm DFB-PMC device spectrum from DFB(e)(f) and PMC facet(g)(h), (i)-(l) optical spectrum for 1567 nm DFB-PMC device from DFB(i)(j) and PMC facet(k)(l).

For the DFB-PMC-DPS device, the fabrication process is the same as that of the DFB-PMC devices, and will not be discussed in further detail.

## 3. Device measurement

### 3.1 DFB-PMC device

As stated above, for the DFB-PMC device, DFBs with three Bragg wavelengths, i.e., 1540 nm, 1550 nm, and 1565 nm were designed and fabricated. Fig. 5 shows their single mode optical spectrum at specific injection DFB currents ($I_{DFB}$) and 2-D optical spectra as a function of $I_{DFB}$ from the DFB rear side and PMC side respectively. The spectrum was measured with a resolution bandwidth (RBW) of 0.06 nm. The measurement results of wavelength, single mode suppression ratios (SMSRs), and average current-induced wavelength redshift coefficient (ACWRC) is listed in Table 1. The measured wavelengths of the DFB lasers with designed Bragg grating wavelengths at 1540 nm, 1550 nm, and 1565 nm are 1543.1 nm, 1555.8 nm, and 1568.1nm, with $I_{DFB}$ set at 150 mA, 170 mA, and 170 mA respectively. Compared to the designed Bragg wavelengths, the measured wavelengths are slightly redshifted due to the heating effect. The corresponding SMSRs are 27 dB, 29 dB, and 38 dB respectively from DFB rear facet. At the PMC output facet, the SMSRs were reduced to 8 dB, 12 dB, and 20 dB. This is due to the significant inter-band and exciton absorption inside the PMC waveguide when the propagating light wavelength is close to the PL wavelength (1530 nm) of the MQW core.

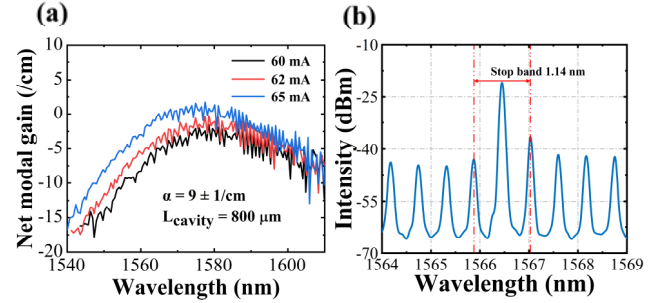

Fig. 6. (a) Measured net modal gain as a function of the wavelength using Haki-Paoli method, (b) optical spectrum of a 600 μm length DFB LD at threshold (48 mA).

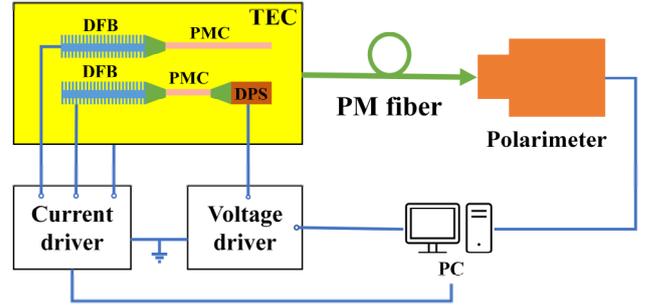

Fig. 7. Experimental setup for the SOP measurement for DFB-PMC and DFB-PMC-DPS devices.

The measured ACWRC from the DFB and PMC sides are 0.025 nm/mA, 0.0207 nm/mA, and 0.0271 nm/mA for Bragg grating wavelengths at 1540 nm, 1550 nm and 1565 nm respectively, all exhibiting stable SLM operation. To estimate the internal loss in the PMC waveguide, an 800 μm length FP laser was fabricated in the same wafer and same fabrication run, and the internal loss in the waveguide was 9/cm measured by the Haki-Paoli method (shown in Fig.6(a)). In order to estimate the κ value of the fabricated DFB lasers, a 600-μm-long DFB laser with a Bragg wavelength of 1565 nm was also fabricated. Fig.6(b) shows the DFB laser's optical spectrum at the threshold current (48 mA). The measured central wavelength is 1566.5 nm and the stop band between two side modes ($\Delta\lambda_s$) is 1.14 nm. The grating coupling coefficient can be estimated using the following formula [17]:

$$\kappa = n_{eff} \frac{\Delta \lambda_s}{\lambda_B^2} \qquad (3)$$

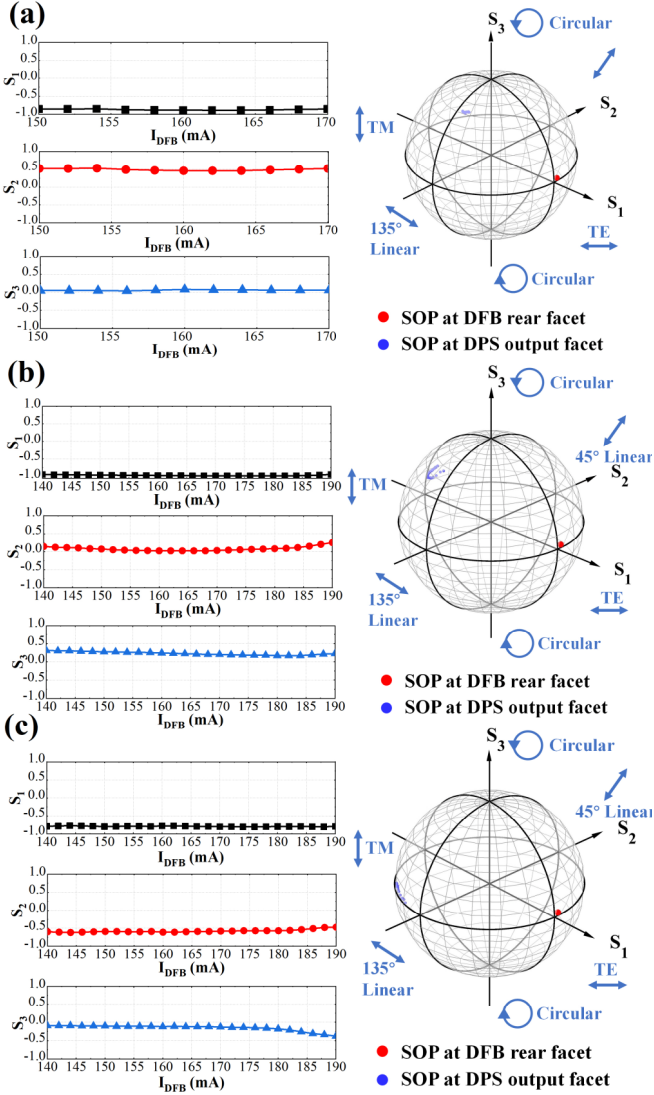

Fig. 8. Measured *SV* at PMC side for (a) 1543 nm, (b) 1555 nm, (c) 1568 nm wavelength DFB-PMC devices.

where $n_{eff}$ is the effective index (3.216); $\lambda_B$ is the lasing wavelength of the DFB laser. The $\kappa$ of the fabricated grating is estimated to be ~15 cm$^{-1}$. $\kappa L$ = 1.8 for the 1200 μm length DFB laser, which ensures stable SLM operation.

The setup of the SOP measurement is shown in Fig. 7. The devices were mounted on a thermoelectric cooler (TEC) and temperature controlled at 20°C as stated previously. The output light from PMC was coupled to a lensed polarization maintaining (PM) fiber and transmitted to a polarimeter to measure the SOP. Both the current driver and the polarimeter were controlled by a computer through the general-purpose interface bus (GPIB) interface by LabVIEW software. We first measured the SOP at the DFB laser rear facet at $I_{DFB}$ from 104 mA to 210 mA, and *S* was constant at (0.998,0.05,0.04). Fig. 8(a)-(c) presents the *SV* at the PMC facets of the DFB-PMC devices with different Bragg wavelengths and the PCE was calculated from Eq. (2) and is listed in Table (2). For the DFB-PMC device with a designed Bragg wavelength at 1550 nm, the average $S_1$ parameter was –0.968 representing a 98.4% PCE for 140 mA < $I_{DFB}$ < 190 mA (corresponding wavelength range 1555.8 – 1556.8 nm). The maximum PCE was 99.1% measured at $I_{DFB}$ =180 mA. For the DFB-PMC device with a designed Bragg wavelength at 1540 nm, the average PCE over the wavelength range from 1543.1 nm to 1544.2 nm was found to be 93% for 150 mA < $I_{DFB}$ < 170 mA, and the maximum PCE was found to be 94.3% at $I_{DFB}$ =162 mA. For the DFB-PMC device with a designed Bragg wavelength at 1565 nm, the average PCE was 89.8% over the range 140 mA < $I_{DFB}$ < 190 mA and wavelength from 1568.1 to 1569.5 nm. The maximum PCE was 90.5% (at $I_{DFB}$ =174 mA). Fig. 9 presents a comparison of the calculated PCE from Full-Wave simulation with the measured average PCE as a function of wavelength. There is very good agreement between the simulation and measurement.

Table 2. Measured PCE from PMC side of the DFB-PMC devices

| Designed DFB laser Wavelengths (nm) | Measured DFB laser Wavelengths (nm) | Current range (mA) | Average PCE | MAX PCE |
|---|---|---|---|---|
| 1540 | 1543.1-1544.2 | 150-170 | 0.930 | 0.943 |
| 1550 | 1555.8-1556.8 | 140-190 | 0.984 | 0.991 |
| 1565 | 1568.1-1569.5 | 140-190 | 0.898 | 0.905 |

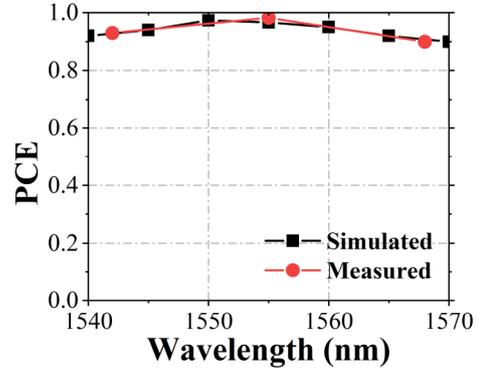

Fig. 9. Simulated and measured PCE versus DFB lasing wavelength.

3.2 DFB-PMC-DPS device

In light of the performance of the DFB-PMC devices, the DFB-PMC-DPS devices were fabricated with a Bragg grating wavelength at 1550 nm. Fig. 10 shows the optical spectrum from the rear side of the DFB section and from the DPS output section where the bias voltage of DPS ($V_{DPS}$) is 0 V. The peak lasing wavelength is at 1557.9 nm with an SMSR of 35 dB at DFB rear facet with $I_{DFB}$ =170 mA, and the ACWRC is 0.023 nm/mA. We first measured the *SV* at the DPS output facet with $V_{DPS}$ =0 V, as shown in Fig. 11(a). The *SV* lies near (0.05, 0.5, 0.865) for the range of $I_{DFB}$ from 160 mA to 175 mA. Then $I_{DFB}$ is fixed at 170 mA and $V_{DPS}$ was gradually increased. The *SV* measured at the DPS output facet is depicted in Fig. 11(b). It is found that the *SV* rotates along the $S_2$-$S_3$ plane, and the rotation angle is about π/3 as $V_{DPS}$ is changed from 0.0 V to –4.0 V in steps of –0.2 V. The rotation angle is limited because the DPS is only 100 μm long in order to reduce the absorption loss inside the DPS.

We note the fabricated PMC length should be kept as close as possible to the designed value by precise control of the cleaving.

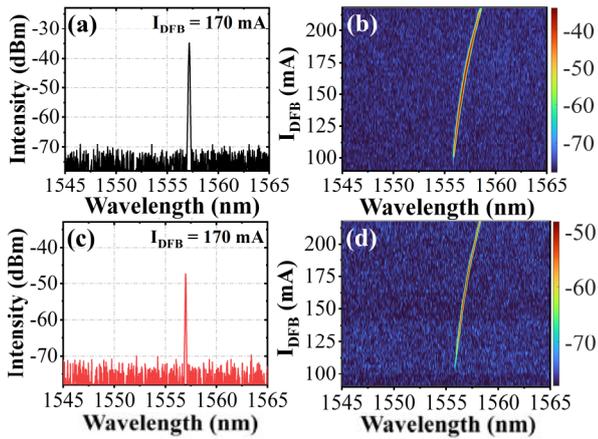

Fig. 10. (a)-(b) Measured optical spectrum from DFB LD rear facet, and (c)-(d) DPS section output facet.

Here a LOOMIS LSD-100 cleaving tool was used with a cleaving accuracy of ±1 μm. The resulting variation in the PCE is less than 0.1%, confirming the tool meets the required tolerance. We also comment that, in order to increase the output power of the device, quantum well intermixing (QWI) could be used to blueshift the bandgap in the PMC and DPS sections and reduce their absorption loss.

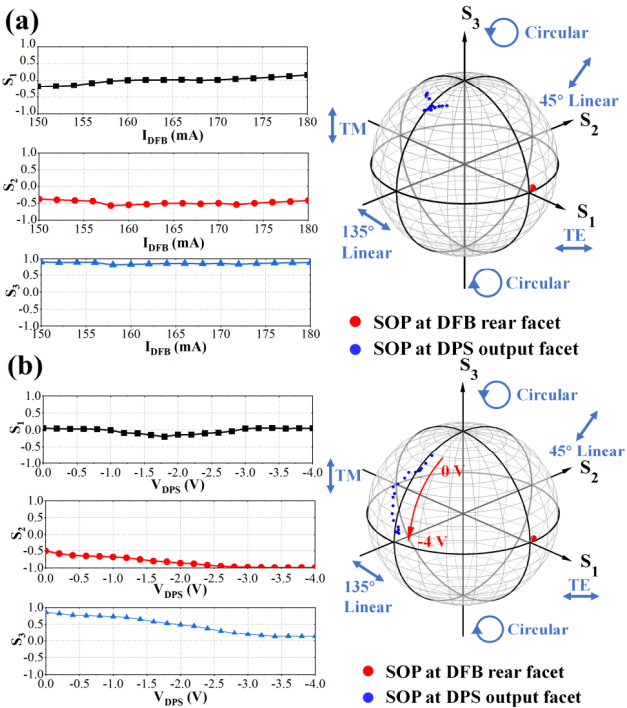

Fig. 11. (a) Measurement $SV$ at DPS side as a function of $I_{DFB}$ with $V_{DPS}$ = 0 V, (b) rotation of $SV$ at DPS output facet as a function of $V_{DPS}$ from DFB-PMC-DPS devices with $I_{DFB}$ = 170 mA.

## 4. Conclusion

We have proposed and fabricated, for the first time, an SWG DFB laser monolithically integrated with a PMC, and an SWG DFB laser monolithically integrated with PMC and DPS based on the IEL PIC scheme. With a 490-μm-long PMC device, a TE/TM conversion efficiency of 98.4 % was obtained over a broad $I_{DFB}$ range from 140 mA to 190 mA at a 1555 nm operating wavelength. Also, the wavelength dependence was calculated and measured; the devices show a >90% PCE over a wavelength range of 1543 nm to 1568 nm. Using the fabricated DFB-PMC-DPS device, a π/3 rotation of the $SV$ was obtained on the surface of the Poincaré sphere over a range of bias voltage from 0 V to −4 V at $I_{DFB}$ =170 mA. A major advantage of the design is that only a single MOVPE step and two dry-etch steps are required to fabricate the device, significantly reducing complexity and cost. The demonstrated devices should be useful in realizing various types of polarization- multiplexed coherent transceivers.

**References**

1. P. J. Winzer, D. T. Neilson, and A. R. Chraplyvy, "Fiber-optic transmission and networking: the previous 20 and the next 20 years [Invited]," Opt. Exp. **26**(18), 24190-24239 (2018).
2. S. Ghosh, Y. Kawabata, T. Tanemura, and Y. Nakano, "Polarization-analyzing circuit on InP for integrated Stokes vector receiver," Opt. Exp. **25**(11), 12303-12310 (2017).
3. B. Holmes, M. Naeem, D. Hutchings, J. Marsh, and A. Kelly, "A semiconductor laser with monolithically integrated dynamic polarization control," Opt. Exp. **20**(18), 20545-20550 (2012).
4. M. Kazi, S. Ghosh, H. Sodabanlu, K. Suzuki, M. Sugiyama, T. Tanemura, and Y. Nakano, "High-speed carrier-injection-based polarization controller with InGaAlAs/InAlAs multiple-quantum wells," IEEE Photon. Technol. Lett. **29**(22), 1951-1954 (2017).
5. S. I. H. Azzam, N. F. Areed, M. M. Abd-Elrazzak, H. El-Mikati, and S. S. Obayya, "Compact polarization rotator based on SOI platform," in *2014 31st National Radio Science Conference (NRSC)*(IEEE2014), pp. 288-293.
6. U. Khalique, Y. Zhu, J. van der Tol, L. Augustin, R. Hanfoug, F. Groen, P. van Veldhoven, M. Smit, M. van de Moosdijk, and W. de Laat, "Ultrashort polarization converter on InP/InGaAsP fabricated by optical lithography," in *Integrated Photonics Research and Applications*(Optica Publishing Group2005), p. IWA3.
7. A. E. Elfiqi, R. Kobayashi, R. Tanomura, T. Tanemura, and Y. Nakano, "Fabrication-tolerant half-ridge InP/InGaAsP polarization rotator with etching-stop layer," IEEE Photon. Technol. Lett. **32**(11), 663-666 (2020).
8. M. Ito, K. Okawa, T. Suganuma, T. Fukui, E. Kato, T. Tanemura, and Y. Nakano, "Efficient InGaAsP MQW-based polarization controller without active-passive integration," Opt. Exp. **29**(7), 10538-10545 (2021).
9. M. A. Naeem, and K. Abid, "A novel full polarisation controller integrated monolithically with a semiconductor laser," in *Asia Communications and Photonics Conference*(Optica Publishing Group2015), p. AM1A. 2.
10. J. Bregenzer, S. McMaster, M. Sorel, B. Holmes, and D. Hutchings, "Polarisation mode converter monolithically integrated within a semiconductor laser," presented at the Conference on Lasers and Electro-Optics2008.
11. A. E. Elfiqi, R. Tanomura, D. Yu, W. Yanwachirakul, H. Shao, Y. Suzuki, T. Tanemura, and Y. Nakano, "Robust InP/InGaAsP polarization rotator based on mode evolution," IEEE Photon. Technol. Lett. **34**(2), 109-112 (2022).
12. X. Sun, W. Cheng, Y. Sun, S. Ye, A. Al-Moathin, Y. Huang, R. Zhang, S. Liang, B. Qiu, and J. Xiong, "Simulation of an AlGaInAs/InP Electro-Absorption Modulator Monolithically Integrated with Sidewall Grating Distributed Feedback Laser by Quantum Well Intermixing," Photonics **9**(8), 564 (2022).
13. X. Sun, S. Ye, W. Cheng, S. Liang, Y. Huang, B. Qiu, Z. Li, J. Xiong, X. Liu, and J. H. J. I. P. J. Marsh, "Monolithically Integrated AlGaInAs MQW Polarization Mode Converter using a Stepped Height Ridge Waveguide," IEEE Photonics Journal **14**(3) (2022).

14. T. Tanemura, and Y. Nakano, "Compact InP Stokes-vector modulator and receiver circuits for short-reach direct-detection optical links," IEICE Trans. Electron. **101**(7), 594-601 (2018).
15. H. Deng, D. O. Yevick, C. Brooks, and P. E. Jessop, "Design Rules for Slanted-Angle Polarization Rotators," J. Light. Technol. **23**(1), 432-445 (2005).
16. L. Hou, M. Tan, M. Haji, I. Eddie, and J. H. Marsh, "EML based on side-wall grating and identical epitaxial layer scheme," IEEE Photon. Technol. Lett. **25**(12), 1169-1172 (2013).
17. M. Razeghi, R. Blondeau, M. Krakowski, J.-C. Bouley, M. Papuchon, B. Cremoux, and J. Duchemin, "Low-threshold distributed feedback lasers fabricated on material grown completely by LP-MOCVD," IEEE J. Quantum Electron **21**(6), 507-511 (1985).